\documentclass[aps,pre,twocolumn,epsfig,floatfix,twoside,a4paper,showpacs]{revtex4-1}
\usepackage{epsfig}

\begin{document} 
\input epsf 
\title{Effect of hydrophobic solutes on the liquid-liquid critical point} 
\author{Dario Corradini,~$^1$ Sergey V. Buldyrev,~$^2$ Paola Gallo,~$^1$ and H. Eugene Stanley~$^3$}
\affiliation{$^1$~Dipartimento di Fisica, Universit\`a Roma Tre, 
Via della Vasca Navale 84, I-00146 Roma, Italy\\
$^2$~Department of Physics, Yeshiva University, 500 West 185th Street, New York, New York 10033, USA\\
$^3$~Center for Polymer Studies and Department of Physics, Boston University, Boston, 
Massachusetts 02215, USA}

\begin{abstract}
  \noindent {Jagla ramp particles, interacting through a ramp
    potential with two characteristic length scales, are known to show
    in their bulk phase thermodynamic and dynamic anomalies, similar
    to what is found in water.  Jagla particles also exhibit a line of
    phase transitions separating a low density liquid phase and a high
    density liquid phase, terminating in a liquid-liquid critical
    point in a region of the phase diagram that can be studied by
    simulations.  Employing molecular dynamics computer simulations,
    we study the thermodynamics and the dynamics of solutions of hard
    spheres (HS) in a solvent formed by Jagla ramp particles.  We
    consider the cases of HS mole fraction $x_{HS}=0.10,0.15$ and
    $0.20$, and also the case $x_{HS}=0.50$ (a 1:1 mixture of HS and
    Jagla particles).  We find a liquid-liquid critical point, up to the highest HS 
    mole fraction; its position shifts to higher pressures and lower temperatures
    upon increasing $x_{HS}$.  We also find that the diffusion
    coefficient anomalies appear to be preserved for all the mole
    fractions studied.}

\end{abstract}



\maketitle

\section{Introduction}\label{intro}

Liquid water exhibits highly unusual thermodynamic and dynamic
behavior~\cite{debenedetti,phystoday,errington}. Among its most known
anomalies there are the decrease in density upon isobaric cooling
(density anomaly), the apparent divergences of thermodynamic response
functions such as the isothermal compressibility, the coefficient of
thermal expansion and the isobaric specific heat upon cooling and the
increase in diffusivity upon isothermal compression (``diffusion
anomaly''). It has been hypothesized that the thermodynamic anomalies
of water may be related to the presence of liquid-liquid (LL) critical
point (LLCP) in the deeply supercooled
region~\cite{poole,poole2,poole3,paschek,paschek2,tanaka,brovchenko,vallauri,mossa,sciortino,sciortino2,mishima,harrington}.
This hypothesized LLCP is the end point
of the liquid-liquid coexistence line separating two distinct liquid
phases: a low density liquid (LDL) and a high density liquid (HDL).

Molecular dynamics simulations of water aiming to study the LLCP and
phenomena related to it often employ water models that reproduce the
tetrahedral orientation-dependent interactions of real water. However
several papers have recently shown that tetrahedrality and even
orientation-dependent interactions are not necessary conditions for
the appearance of the density and the diffusion anomalies or the
presence of a LL
transition~\cite{Canpolat98,Sadr98,xu,xu2,xu3,xu4,kumar,yan,gibson,lomba}. 
There exists a family of spherically symmetric potentials composed by a hard
core and a linear repulsive ramp, the Jagla ramp
model~\cite{jagla,jagla2} that can be tuned, varying the ratio between its
two characteristic length scales, in order to span the range of behavior
from hard spheres to water-like~\cite{yan,yan2,yan3}. With the
appropriate choice of parameters the Jagla ramp potential with two
characteristic length scales  displays both thermodynamic and dynamic anomalies
and a LL transition.

Spherically symmetric potentials with softened core have been used as coarse-grained 
models for a variety of substances beside water, such as metallic systems and
colloidal suspensions~\cite{fornleitner,malescio,collab}.

Besides its rather unique behavior as a pure liquid, water is also a
remarkable solvent. In particular, phenomena related to the solvation
of apolar solutes in water are interesting since they encompass
biological membrane formation, globular protein folding and also the
stability of mesoscopic
assembly~\cite{sergey,chandler,tanford,kauzmann}.  A large number of
papers have in the past addressed the phenomenon of hydrophobic
hydration (see for example
Refs.~\onlinecite{paschek,chandler,ashbaugh,ashbaugh2,ashbaugh3,stillinger,pratt,pratt2,garde,hummer,hummer2,
  rajamani,widom,holzmann}).  Small apolar solutes, such as alkanes or
noble gases are poorly soluble in water. The solvation free energy of
this kind of solutes is large and positive due to the large and
negative entropy contribution, the latter having been related to the
structure of the hydrophobic hydration shell~\cite{widom}. These
quantities have a marked temperature dependence and one intriguing
anomaly of water in solutions is represented by the increase in
solubility of hydrophobic gases upon decreasing
temperature~\cite{wilhelm}.

\begin{figure}[h!] \centerline{\psfig{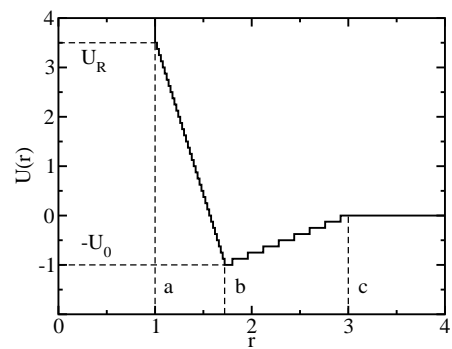}}
\caption{Spherically symmetric Jagla ramp potential. The potential has
two scales: the hard core diameter $r=a$ and the soft-core diameter
$r=b$. In this case $U_R/U_0=3.56$, $b/a=1.72$ and $c/a=3$. We have
discretized the potential using a discretization step $\Delta
U$=$U_0/8$.}
\label{fig:1}
\end{figure}

\begin{figure*}[ht] \centerline{\psfig{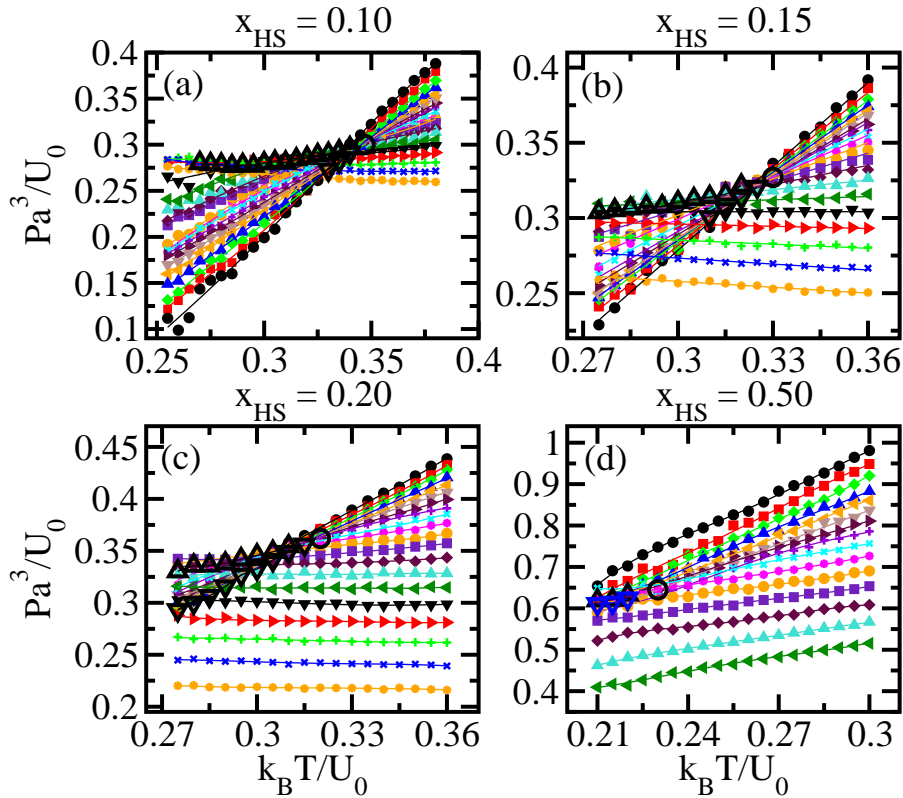}}
\caption{(Color online) Isochores in the $P-T$ plane for the four
solutions at different hard spheres mole fractions. (a) $x_{HS}=0.10$,
isochores are drawn for $\rho\equiv Na^3/L^3$ where
$L/a=17.4,17.3,\dots,15.5$.  The range of corresponding densities span
from $\rho=0.328$ to $\rho=0.464$ (from bottom to top).  The
temperatures range is $0.255\leq T \leq 0.380$. (b) $x_{HS}=0.15$,
isochores are drawn for $\rho\equiv Na^3/L^3$ where
$L/a=17.4,17.3,\dots,15.5$. The range of corresponding densities span
from $\rho=0.328$ to $\rho=0.464$ (from bottom to top). The
temperatures range is $0.275\leq T \leq 0.360$.  (c) $x_{HS}=0.20$,
isochores are drawn for $\rho\equiv Na^3/L^3$ where
$L/a=17.4,17.3,\dots,15.5$.  The range of corresponding densities span
from $\rho=0.328$ to $\rho=0.464$ (from bottom to top).  The
temperatures range is $0.275\leq T \leq 0.360$. (d) $x_{HS}=0.50$,
isochores are drawn for densities $\rho\equiv Na^3/L^3$ where
$L/a=15.1,15,0,\dots,13.7$.  The range of corresponding densities span
from $\rho=0.502$ to $\rho=0.672$ (from bottom to top). The
temperatures range is $0.210\leq T \leq 0.300$.  In all panels the
lines are fourth degree polynomial fits to simulated state points. For
all mole fractions the position of the LLCP (circles), the LDL LMS
(triangles up) and the HDL LMS (triangles down) are also reported.}
\label{fig:2}
\end{figure*}

Experimentally the solubility of small apolar hydrophobic solutes
decreases upon decreasing temperature until a minimum is reached in
the temperature range from 310~K to 350~K. Upon further temperature
decrease, the solubility increase
monotonically~\cite{ashbaugh,sergey}.  
The solubility of model
hydrophobic solutes, hard spheres, in the two-scale ramp
potential particles has been recently assessed~\cite{sergey}. It was
found that the mixture of ramp particles and hard spheres shows a
temperature of minimum solubility, similarly to experimental results
in water. The increased solubility upon cooling is connected to
the presence of two repulsive length scales in the model potential, the
hard core corresponding to nearest-neighbor shell of solvent molecules
and the soft repulsive core. Moreover it was observed that hard
spheres are more favorably solvated in low density phases, in
accord with what found in simulations of water~\cite{paschek}.

The nature of critical phenomena in the presence of solutes has been
extensively studied in literature with regard to the liquid-gas
critical point~\cite{kony,scott,abdul}, while the effect of solutes on
the LLCP of the solvent is a relatively new
subject~\cite{chatterjee,corradini10}.
In this work we investigate the thermodynamic and dynamic properties of 
the mixture of ramp potential supplemented by an attractive tail 
and apolar solutes modeled by hard spheres. We examine three mole fractions
of hard spheres, $x_{HS}=0.10,0.15$ and $0.20$.
We also investigate thermodynamics and diffusivity for a 1:1 mixture of hard spheres 
and Jagla ramp particles, $x_{HS}=0.50$.
The paper is structured as follows. In Sec.~\ref{methods} we give the
details of the interaction potential and of the simulation method. 
We present results and discussion in Sec.~\ref{results}, and conclusions
 in Sec.~\ref{conclu}
 
\begin{figure*}[ht] \centerline{\psfig{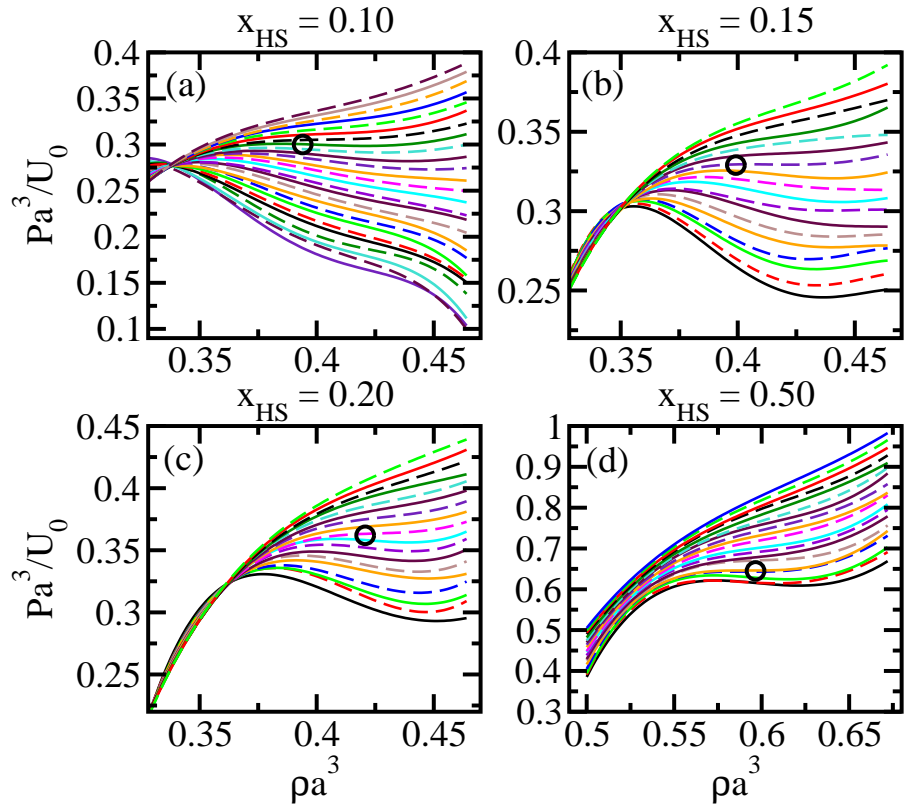}}
\caption{(Color online) Isotherms in the $P-\rho$ plane for the four
solutions at different hard spheres mole fractions. (a) $x_{HS}=0.10$,
isotherms are drawn from $T=0.255$ to $T=0.380$ every $\Delta T=0.005$
(from bottom to top) . (b) $x_{HS}=0.15$, isotherms are drawn from
$T=0.275$ to $T=0.360$ every $\Delta T=0.005$ (from bottom to
top). (c) $x_{HS}=0.20$, isotherms are drawn from $T=0.275$ to
$T=0.360$ every $\Delta T=0.005$ (from bottom to top). (d)
$x_{HS}=0.50$, isotherms are drawn from $T=0.210$ to $T=0.300$ every
$\Delta T=0.005$ (from bottom to top). In all panels the lines are
fourth degree polynomial fits to simulated state points. The circles
represent the position of the LLCP. Every other line has been made
dashed to help distinguishing in between them.}
\label{fig:3}
\end{figure*}

\section{Methods}\label{methods}

We perform discrete molecular dynamics~\cite{kumar,xu3,buldyrev} on
systems composed by $N=N_{ramp}+N_{HS}=1728$ particle, where
$N_{ramp}$ is the number of water-like particles and $N_{HS}$ is the
number of hard spheres. We study four systems with different
composition.  The solute content in the four systems is
$N_{HS}=173$ for $x_{HS}=0.10$, $N_{HS}=260$ for $x_{HS}=0.15$,
$N_{HS}=345$ for $x_{HS}=0.20$ and $N_{HS}=864$ for $x_{HS}=0.50$

The pair-wise Jagla ramp interaction potential~\cite{jagla,jagla2} has
two characteristic length scales, the hard-core distance $r=a$ and the
soft-core distance $r=b$. The minimum of the energy $U_0$ is
corresponds to the soft-core distance. An attractive tail extends up
to $r=c$. The potential has been discretized, in order to be able to
employ the algorithm of discrete molecular dynamics. We partition the repulsive
ramp  into 36 steps of width $0.02a$ and the
attractive ramp into eight steps of width $0.16a$. The $\Delta U$ at each
step is $U_0/8=0.125$.  The parameters of the ramp
potential~\cite{xu3} have been set to $b/a=1.72$ and
$c/a=3$. $U_R=3.56\, U_0$ is defined as the value of the energy at
$r=a$, obtained via least-squares linear fit to the discretized
repulsive ramp.  The shape of the spherically symmetric Jagla ramp
potential employed in this work is shown in Fig.~\ref{fig:1}. As
in previous papers~\cite{xu2,xu3}, this parametrization of the
ramp potential prevents the occurrence of crystallization. Furthermore, with this
choice of parameters, the line of LL phase coexistence extends into
the equilibrium liquid phase and  ends in a LLCP.  The diameter of
the hard spheres is $a$, the same as the hard-core distance of Jagla
ramp interaction potential. The solvent and the solute interact via
a hard-core potential.

We express all quantities in reduced units. Distances are in unit of
$a$, energies in units of $U_0$ and time in units $a\sqrt{m/U_0}$,
where $m$ is the mass that is assumed to be unitary. The density
defined as $\rho\equiv {N}/{L^3}$, where $L$ is the edge of the cubic
simulation box, is measured in units of $a^{-3}$, pressure in unit of
$U_0/a^3$ and temperature in units of $U_0/k_B$.

We perform simulations at constant $N$, $V$ and $T$, with $T$
controlled by rescaling the velocity of the particles with a modified
Berendsen algorithm (details are given in Ref.~\onlinecite{buldyrev}), or
at constant $N$, $P$ and $T$, with $P$ controlled by allowing the edge
of the simulation box to vary with time applying the standard
Berendsen algorithm~\cite{buldyrev}.

\begin{figure}[ht] \centerline{\psfig{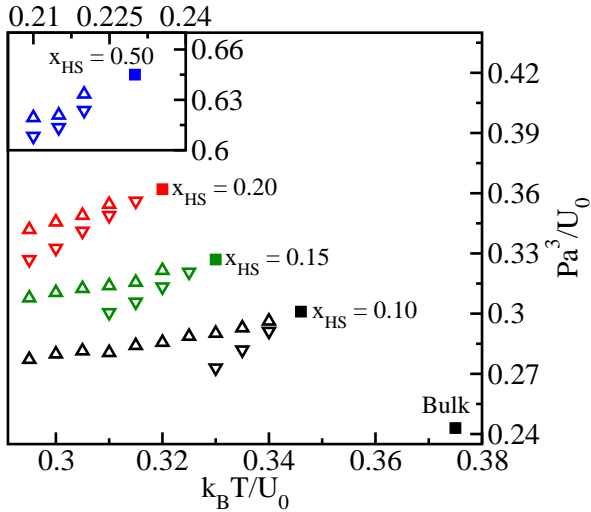}}
\caption{Position in the $P-T$ plane of the LLCP (squares) and of LDL
(triangles up) and HDL (triangles down) LMS lines for the solutions of
hard spheres in Jagla ramp particles, with $x_{HS}=0.10, 0.15$ and
$0.20$. The position of the LLCP of bulk Jagla ramp particles is also
reported for comparison. In the inset the analogous quantities are
shown for the 1:1 mixture of Jagla ramp particles and hard spheres.}
\label{fig:4}
\end{figure}

\section{Results and Discussion}\label{results}

We study the thermodynamics of the solutions of hard spheres in Jagla
ramp potential water-like particles, analyzing the isochores in the
$P-T$ plane and the isotherms in the $P-\rho$ plane.  We study ranges
of temperature and density for which the hard spheres are completely
soluble in the Jagla ramp potential liquid~\cite{sergey}. As a
consequence in the range we study here, no solvent-solute
demixing occurs.  The position of the LLCP has been estimated
considering the inflection point in the isotherms where
\begin{equation}
\left(\frac{\partial P}{\partial \rho} \right)_T=\left(\frac{\partial^2 P}{\partial \rho^2} \right)_T=0 
\end{equation}
The points of the LL limit of mechanical stability (LMS) lines have
been determined by the points for which $(\partial P/\partial
\rho)_T=0$, and $(\partial^2 P/\partial \rho^2)_T\neq 0$ where the
isothermal compressibility $K_T =(\partial \rho/\partial P)_T/\rho$
diverges.
 
The thermodynamic properties of the bulk Jagla ramp potential
particles, with the same set of parameters used in this work, have
been previously assessed~\cite{xu2}. In particular the coordinates in
the thermodynamic plane of the LLCP of bulk Jagla ramp potential
particles are $T_c=0.375$, $P_c=0.243$ and $\rho_c=0.37$.

In Fig.~\ref{fig:2} we present the isochores $P-T$ plane for all the
solutions at different mole fractions.  In this figure the location of
the LLCP and the two branches of the LL LMS lines are also shown.  
For mole fractions up to $x_{HS}=0.20$
we observe that the isochores converge very clearly to the LLCP and
cross at points corresponding to the LMS lines. For mole fractions
$x_{HS}=0.10,0.15$ and $0.20$ we also note that the low density
isochores appear almost flat, signaling the presence of the density
anomaly. The points of the temperature of maximum density line are in
fact given by the zeros of the coefficient of thermal expansion, where
$(\partial P/\partial T)_\rho=0$.  In the case of the mixture with
$x_{HS}=0.50$ the convergence of the isochores to the LLCP appears
less precise. In this case we observe that the isochores are not flat
any longer, indicating the disappearance of the density anomaly at
this high mole fraction of hard spheres.

Fig.~\ref{fig:3} shows the isotherms of the mixtures with
$x_{HS}=0.10,0.15,0.20$ and $0.50$  along with the
position of the LLCP in the $P-\rho$ plane.  We can observe the
inflection point in the isotherms corresponding to the critical point
below which the isotherms show van der Waals-like loops indicating LL
coexistence. We also observe that upon increasing the mole fraction of
solutes the density range of the region of coexistence narrows. We
also point out that for low density the isotherms cross due to density
anomaly for mole fractions $x_{HS}=0.10,0.15$ and $0.20$ .  In fact
for the points where isotherms cross, $(\partial
P/\partial T)_\rho=0$, so the crossing of the isotherms indicates a 
density anomaly.  Also in the case of $x_{HS}=0.50$ mixture the
presence of a LLCP can be highlighted by the inflection point in the
critical isotherm and the van der Waals-like loop structure of the
isotherms below the critical temperature. An important feature of
isotherms of this system is that they do not cross, confirming what we
have found looking at the isochores in Fig.~\ref{fig:2}. For this very
high mole fraction of solutes the density anomaly is not longer present,
while the LLCP is still found.

\begin{table}[t]
\label{table:1}
\begin{center}
\caption{Position of the LLCP for bulk Jagla ramp potential particles
and for the solutions of hard spheres in Jagla ramp potential
particles with $x_{HS}=0.10,0.15$ and $0.20$ and for 1:1 mixture of
Jagla ramp potential particles and hard spheres ($x_{HS}=0.50$).}
\begin{ruledtabular}
\begin{tabular}{cccc} $x_{HS}$&$T_c$& $P_c$&$\rho_c$\\
\hline
0.00 (bulk)&0.375&0.243&0.370\\
0.10&0.346&0.301&0.394\\
0.15&0.330&0.327&0.399\\
0.20&0.320&0.362&0.421\\
0.50 (1:1 mixture)&0.230&0.645&0.597 
\end{tabular}
\end{ruledtabular}
\end{center}
\end{table}

The coordinates of the estimated positions of the LLCP for bulk Jagla
ramp potential particles~\cite{xu2} and for all the systems with
different composition are reported together in Table I.

\begin{figure*}[t!] \centerline{\psfig{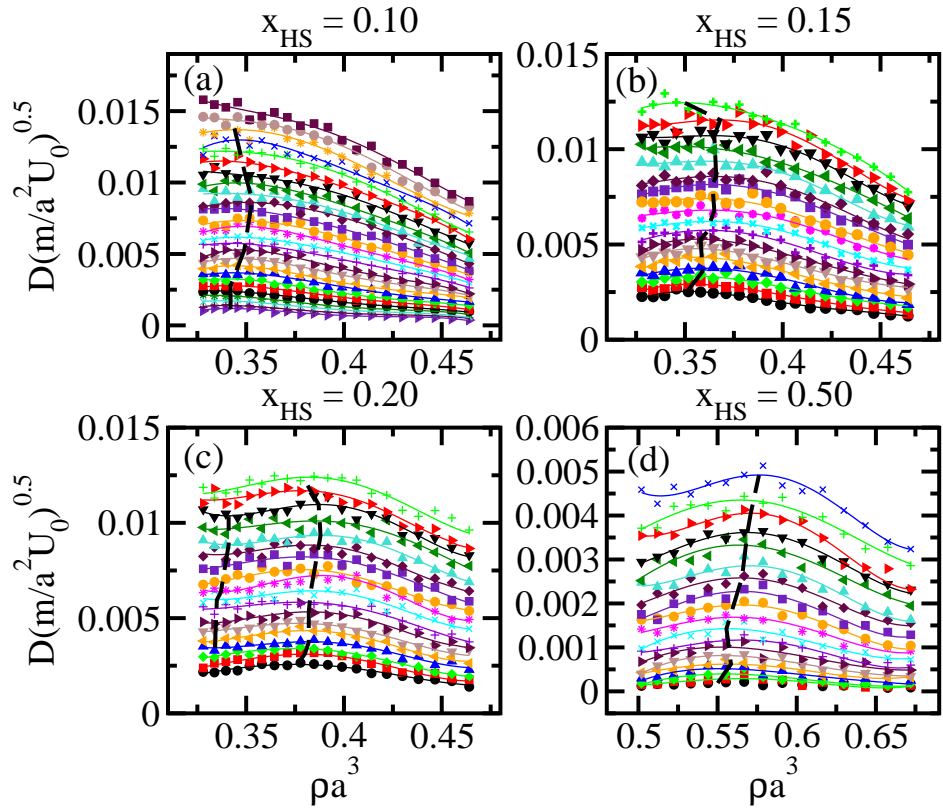}}
\caption{(Color online) Diffusion coefficients of Jagla ramp particles at constant
temperature for the four solutions at different hard spheres mole
fractions. (a) $x_{HS}=0.10$, isotherms of the diffusion coefficient
are drawn from $T=0.255$ to $T=0.380$ every $\Delta T=0.005$ (from
bottom to top). (b) $x_{HS}=0.15$, isotherms of the diffusion
coefficient are drawn from $T=0.275$ to $T=0.360$ every $\Delta
T=0.005$ (from bottom to top). (c) $x_{HS}=0.20$, isotherms of the
diffusion coefficient are drawn from $T=0.275$ to $T=0.360$ every
$\Delta T=0.005$ (from bottom to top). (d) $x_{HS}=0.50$, isotherms of
the diffusion coefficient are drawn from $T=0.210$ to $T=0.300$ every
$\Delta T=0.005$ (from bottom to top). For all panels, the lines are
fourth degree polynomial fits to simulated state points. The dashed
lines join the diffusivity extrema (maxima for
$x_{HS}=0.10,0.15,0.50$, maxima and minima for $x_{HS}=0.20$).}
\label{fig:5}
\end{figure*}

\begin{figure}[t] \centerline{\psfig{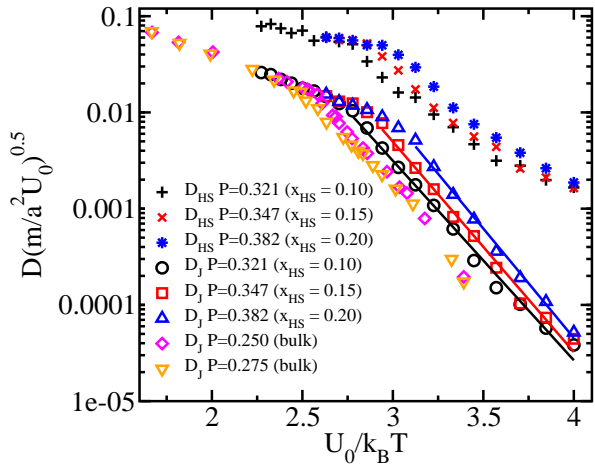}}
\caption{Diffusion coefficients for hard spheres and Jagla ramp
particles along a constant pressure path above the critical
point. $P=0.321$ for $x_{HS}=0.10$, $P=0.347$ for $x_{HS}=0.15$ and
$P=0.382$ for $x_{HS}=0.20$.  Lines correspond to the Arrhenius fit
for the Jagla ramp particles. In this figure the diffusion
coefficients at constant pressure for the bulk Jagla ramp particles at
$P=0.250$ and $P=0.275$ are also reported for comparison.}
\label{fig:6}
\end{figure}

\begin{figure*}[t] 
\centerline{\psfig{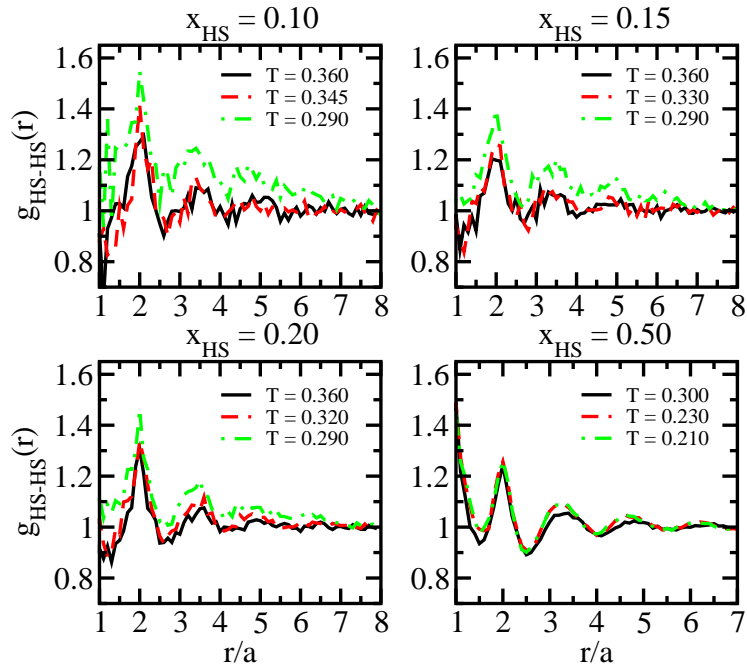}}
\caption{Hard sphere-hard sphere radial distribution functions for the
mixtures with $x_{HS}=0.10,0.15,0.20$ and $0.50$ calculated at their critical
densities (see Table I). The $g_{HS-HS}(r)$ are reported at temperatures T=0.360,
0.345 and 0.290 for $x_{HS}=0.10$, T=0.360, 0.330 and 0.290 for
$x_{HS}=0.15$, T=0.360, 0.320 and 0.290 for $x_{HS}=0.15$ and T=0.300,
0.230 and 0.210 for $x_{HS}=0.50$.}
\label{fig:7}
\end{figure*}

The positions of the LLCP and the two branches of the LL LMS line for
the solutions with $x_{HS}=0.10,0.15$ and $0.20$ and for the 1:1
system are also reported in Fig.~\ref{fig:4} in the $P-T$ plane. The
position of the LLCP moves to higher pressures and lower temperatures
upon increasing the solute mole fraction. This shift of the critical
point could be connected to the fact that hard spheres are more
favorably solvated in LDL~\cite{paschek,sergey}. In fact the LDL
region of the phase diagram progressively widens upon increasing the
mole fraction of hard spheres.

Looking at the isochores (Fig.~\ref{fig:2}) and at the isotherms
(Fig.~\ref{fig:3}) we can observe that upon increasing the mole
fraction of hard spheres, the width of the coexistence envelope is
reduced. In fact the region of crossing of the isochores is
progressively reduced upon increasing the solute mole fraction, on
going from the system at $x_{HS}=0.10$ to the one at $x_{HS}=0.50$.
Also the width of the loop region in the isotherms plane becomes more
narrow spanning a minor range of densities, upon increasing the solute
mole fraction.  From these considerations we can argue that upon a
further increase in solute mole fraction, the LL critical phenomenon
will disappear. This is in agreement with studies of the LL transition
in aqueous
solutions~\cite{corradini10,archer,corradini08,corradini09}.

In Fig.~\ref{fig:5} we present the diffusion coefficients for the
Jagla ramp potential particles, calculated at constant temperature as
a function of the density of the system, for all the mole fractions
studied. For all solute mole fractions, we find the appearance of the
diffusivity anomaly.  The $x_{HS}=0.10,0.15$ and $0.20$ mixtures are
studied in the same set of densities, $\rho=0.328$ to $\rho=0.464$.
In this range maxima are evident for $x_{HS}=0.10,0.15$ and $0.20$,
while minima can be seen only for the $x_{HS}=0.20$ mixture. Minima
for the $x_{HS}=0.10$ and $0.15$ mixtures are to be found at lower
densities, out of the spanned density range.  This is due to the
narrowing of the region of the diffusivity anomaly upon increasing the
solute mole fraction.  Thus not only thermodynamic anomalies but also
dynamic anomalies exist in a smaller range of densities, upon
increasing the solute mole fraction.  The diffusion coefficients at
constant pressure for the $x_{HS}=0.50$ are studied in the density
range $\rho=0.502$ to $\rho=0.672$, thus cannot be directly compared
with mixtures with lower solute mole fraction as the range of densities
spanned is different. However we can see that they also exhibit
diffusion anomaly with the presence of maxima in the isotherms of the
diffusion coefficient.

In Fig.~\ref{fig:6} we report the behavior of the diffusion
coefficient of Jagla ramp particles and hard spheres calculated for a
constant pressure path above the critical pressure for the solutions
with hard sphere mole fractions $x_{HS}=0.10, 0.15$ and $0.20$.  For
all compositions the pressure is set to the critical pressure plus
$\Delta P=0.020$.  In the bulk Jagla ramp particles system it was
found that the trend of the diffusion coefficient, calculated on a
cooling path above the critical point, shows a crossover from a high
temperature behavior (LDL-like) to a stronger (HDL-like)
behavior~\cite{xu2}. This change in trend was connected to the maximum
of the specific heat that occurs at the Widom line via the Adam-Gibbs
equation $D=D_0 \exp(-C/TS_{conf})$, where $S_{conf}$ is the configurational
entropy.  We can observe that the crossover in the behavior of the
diffusion coefficient is maintained up to the highest mole fraction
studied ($x_{HS}=0.20$), and at low enough temperature, below the
Widom line, for all compositions the diffusion behavior becomes that
of a HDL Jagla liquid.  The temperature at which the dynamic crossover
occurs decreases upon increasing the mole fraction of solutes. As the
Widom line is connected to the LLCP, the shift to lower temperatures
of the dynamic crossover confirms the shift to lower temperatures of
the LLCP we have found studying the thermodynamics (see
Fig.~\ref{fig:4}).  The trend of the diffusion coefficient of hard
spheres closely follows that of the Jagla ramp potential particles,
thus we can derive that the diffusive behavior of the hard spheres
solute is determined by the solvent.

Therefore we conclude that the dynamic cross-over found in bulk Jagla
ramp particles system (analogous to liquid water) at the Widom line,
is preserved in the solutions and that the temperature of dynamic
cross-over decreases, upon increasing the mole fraction of hydrophobic
solutes.

Finally the solute-solute radial distribution functions are shown in Fig.~\ref{fig:7}. The 
$g_{HS-HS}(r)$ have been reported for the critical density of the systems and for three temperatures, 
T=0.360, the critical temperature and T=0.290 for the solutions with 
$x_{HS}=0.10,0.15$ and $0.20$ and for T=0.300, the critical temperature and T=0.210  for $x_{HS}=0.50$. 
A small tendency of the solutes to cluster upon increasing concentration can be seen looking at the 
progressive increase in the $g(r)$ near the hard-core distance. 
At the lowest temperature the solute-solute radial distribution functions also
show anomalous behavior with a progressive decay toward the asymptotic value of 1 at large $r$ that 
tends to disappear upon increasing the mole fraction of solutes. The decay indicates segregation of
the HS into the LDL phase below the critical point.

\section{Conclusions}\label{conclu}

We performed discrete molecular dynamics simulations on the mixture of Jagla ramp potential 
(water-like) particles and hard spheres, at four different compositions, 
$x_{HS}=0.10,0.15,0.20$ and $0.50$. The thermodynamic and dynamic behavior
were studied for the solutions with $x_{HS}=0.10,0.15$ and $0.20$. Thermodynamics
and diffusive behavior were also studied for the 1:1 mixture of Jagla ramp particles and hard spheres.

The analysis of the isochores and the isotherms plane revealed the
presence of a liquid-liquid critical point for all the investigated
system. We also found that while density anomaly is present in the
solutions with $x_{HS}=0.10,0.15$ and $0.20$, it disappears for the
1:1 mixture.  Furthermore we also observe a narrowing of the
coexistence envelope in both planes, upon increasing the solute mole
fraction.  The position of the critical point is found to shift toward
lower temperatures and higher pressures, upon increasing the hard
sphere mole fraction. This shift may be related to the favored
solvation of hard spheres in the LDL Jagla ramp potential particles
solvent.

The complete phase diagram of the solution in the vicinity
of the LLCP may be quite complex and might include also an upper critical solution
temperature different from the LLCP. The interplay of these two critical points is
an interesting subject which deserves separate investigation.

The appearance of extrema in the behavior of the constant temperature
diffusion coefficient for Jagla ramp potential particles, i.e. the
diffusion anomaly is also preserved up to the highest mole fraction of
hard spheres studied, with a narrowing of the anomalous region upon
increasing the solute mole fraction, also for this dynamic property.

Finally a change in trend of the constant pressure diffusion
coefficient from LDL-like behavior to HDL-like behavior can be
observed when cooling the system in a constant pressure path, above
the critical pressure. This cross-over in the dynamic behavior can be
related to the crossing of the Widom line, above the critical
point. The dynamic cross-over observed for bulk Jagla ramp particles
at the Widom line shifts to lower temperatures upon increasing the
content of solutes.

\section*{ACKNOWLEDGMENTS}
D.~C. and P.~G. gratefully acknowledge the computational support given
by INFN-GRID at University Roma Tre and by the Democritos National
Simulation Center at SISSA. S.V.B. and H.E.S. are supported by the NSF
Chemistry Division.   S.V.B. acknowledges partial support through the
Bernard W. Gamson Computational Science Center at Yeshiva College.

\end{document}